\documentclass[10pt,conference]{IEEEtran}

\usepackage{xspace}
\usepackage{multirow}
\usepackage{multicol}
\usepackage{enumitem}
\usepackage{listings}
\usepackage{siunitx}
\usepackage{booktabs}
\usepackage{booktabs}
\usepackage{longtable}
\usepackage{multirow}
\usepackage{tabularx}
\usepackage{booktabs}
\usepackage{makecell}
\usepackage{tabularx}
\usepackage{listings}
\usepackage[table]{xcolor}
\usepackage{graphicx}
\usepackage{subcaption}
\usepackage{caption}
\usepackage{inconsolata}
\usepackage{array}
\usepackage{amsmath}
\usepackage{algorithm}
\usepackage{algorithmicx}
\usepackage{algpseudocode}
\usepackage[most]{tcolorbox}
\usepackage{hyperref}
\hypersetup{
    colorlinks=true,
    linkcolor=blue,
    filecolor=magenta,      
    urlcolor=cyan,
    pdftitle={Overleaf Example},
    pdfpagemode=FullScreen,
    }

\usepackage{amsthm}
\theoremstyle{definition}
\newtheoremstyle{mydefstyle}
  {}
  {}
  {\normalfont}
  {}
  {\bfseries}
  {\bfseries.}
  {.5em}
  {{\bfseries\thmname{#1}\thmnumber{ #2}}\thmnote{ (#3)}}

\theoremstyle{mydefstyle}

\algnewcommand\algorithmicforeach{\textbf{for each}}
\algdef{S}[FOR]{ForEach}[1]{\algorithmicforeach\ #1\ \algorithmicdo}

\newcommand{\tool}{\textsc{Melt}\xspace}
\newcommand{\comby}{\texttt{Comby}\xspace}
\newcommand{\pandas}{\texttt{pandas}\xspace}
\newcommand{\scipy}{\texttt{scipy}\xspace}
\newcommand{\sklearn}{\texttt{sklearn}\xspace}
\newcommand{\numpy}{\texttt{numpy}\xspace}
\newcommand{\jedi}{\texttt{Jedi}\xspace}
\newcommand{\inferrule}{\texttt{InferRules}\xspace}
\newcommand{\PyEvolve}{\texttt{PyEvolve}\xspace}
\newcommand{\SOAR}{\texttt{SOAR}\xspace}
\newcommand{\gpt}{\texttt{GPT-4}\xspace}

\usepackage{etoolbox} 
\let\oldttfamily\ttfamily 
\renewcommand{\ttfamily}{\oldttfamily\small}

\definecolor{mycolor}{RGB}{111, 111, 111}

\newtcolorbox{highlight}[1][]{
  colback=mycolor!10!white,
  colframe=mycolor!80!black,
  fonttitle=\bfseries,
  coltitle=mycolor!80!black,
  colbacktitle=mycolor!20!white,
  boxrule=1pt,
  arc=3pt,
  outer arc=3pt,
  boxsep=0pt,
  left=10pt,
  right=10pt,
  top=6pt,
  bottom=6pt,
  toptitle=2pt,
  bottomtitle=2pt,
  lefttitle=2pt,
  righttitle=2pt,
  titlerule=0pt,
  attach boxed title to top left={yshift=-\tcboxedtitleheight/2, xshift=2mm},
  boxed title style={arc=3pt, outer arc=3pt},
}

\definecolor{darkblue}{HTML}{1F77B4}
\definecolor{darkorange}{HTML}{FF7F0E}
\definecolor{darkgreen}{HTML}{006400}
\definecolor{darkred}{HTML}{D62728}
\definecolor{darkpurple}{HTML}{9467BD}
\definecolor{darkbrown}{HTML}{8C564B}
\definecolor{darkpink}{HTML}{E377C2}
\definecolor{darkgray}{HTML}{7F7F7F}
\definecolor{darkcyan}{HTML}{17BECF}

\lstdefinelanguage{langs}{
  moredelim=[is][\color{darkgreen}]{*}{*},
  moredelim=*[s][\color{black}]{.}{\ },
  moredelim=*[s][\color{black}]{(}{)},
  morekeywords={},
}

\lstdefinestyle{pythonstyle}{
language=Python,
morekeywords={self, assert, True},
keywordstyle=\color{darkblue},
emph={read_csv},          
emphstyle=\color{blue},    
stringstyle=\color{green},
showstringspaces=false, 
escapechar=@, 
basicstyle=\small\ttfamily\selectfont,
moredelim=[is][\color{darkgreen}]{*}{*},
}

\lstdefinestyle{pythonstyle2}{
language=Python,
morekeywords={self, assert, True, compression, encoding, index_col},
keywordstyle=\color{black},
emph={read_csv, squeeze},          
emphstyle=\color{blue},    
stringstyle=\color{green},
showstringspaces=false, 
escapechar=@, 
basicstyle=\small\ttfamily\selectfont,
moredelim=[is][\color{darkgreen}]{*}{*},
}

\lstnewenvironment{yourcodelisting}[1][]{
  \lstset{language=langs, columns=flexible, basicstyle=\small\ttfamily\selectfont, #1}
}{}

\lstdefinestyle{javastyle} {
  language=Java,
  showspaces=false,
  showtabs=false,
    tabsize=4,
  breaklines=true,
  showstringspaces=false,
  breakatwhitespace=true,
  commentstyle=\color{dkred},
  stringstyle=\color{dkgreen},
  keywordstyle=\color{blue},
  ndkeywordstyle=\color{red},
  basicstyle=\footnotesize\ttfamily,
  numberstyle=\ttfamily\footnotesize\color{gray},
  numbers=left,
  stepnumber=1,    
  firstnumber=1,
  numberfirstline=true,
    numbersep=10pt,
  escapechar=@, 
    xleftmargin=.23in
}

\definecolor{myblue}{RGB}{0, 128, 128}
\definecolor{amethyst}{rgb}{0.6, 0.4, 0.8}

\begin{document}

\title{MELT: Mining Effective Lightweight Transformations from Pull Requests}

\makeatletter
\newcommand{\linebreakand}{%
  \end{@IEEEauthorhalign}
  \hfill\mbox{}\par
  \mbox{}\hfill\begin{@IEEEauthorhalign}
}
\makeatother

\author{
    \IEEEauthorblockN{Daniel Ramos}
    \IEEEauthorblockA{School of Computer Science, INESC-ID\\
                      Carnegie Mellon University, USA\\
                      danielrr@cmu.edu}
    \and
    \IEEEauthorblockN{Hailie Mitchell}
    \IEEEauthorblockA{Computer Science Department\\
                      Dickinson College, USA\\
                      mitchelh@dickinson.edu}
    \and
    \IEEEauthorblockN{Inês Lynce}
    \IEEEauthorblockA{INESC-ID, Instituto Superior Técnico\\
                      Universidade de Lisboa, Portugal\\
                      ines.lynce@tecnico.ulisboa.pt}
    \linebreakand
    \IEEEauthorblockN{Vasco Manquinho}
    \IEEEauthorblockA{INESC-ID, Instituto Superior Técnico\\
                      Universidade de Lisboa, Portugal\\
                      vasco.manquinho@inesc-id.pt}
    \and
    \IEEEauthorblockN{Ruben Martins}
    \IEEEauthorblockA{School of Computer Science\\
                      Carnegie Mellon University, USA\\
                      rubenm@andrew.cmu.edu}
    \and
    \IEEEauthorblockN{Claire Le Goues}
    \IEEEauthorblockA{School of Computer Science\\
                      Carnegie Mellon University, USA\\
                      clegoues@cs.cmu.edu}
}

\maketitle

\thispagestyle{plain}
\pagestyle{plain}

\begin{abstract}

Software developers often struggle to update APIs, leading to manual, time-consuming, and error-prone processes. 
We introduce \tool, a new approach that generates lightweight API migration rules directly from pull requests in popular library repositories.
Our key insight is that pull requests merged into open-source libraries are a rich source of information sufficient to mine API migration rules.
By leveraging code examples mined from the library source and automatically generated code examples based on the pull requests, we infer transformation rules in \comby, a language for structural code search and replace. 
Since inferred rules from single code examples may be too specific, we propose a generalization procedure to make the rules more applicable to client projects. 
\tool rules are syntax-driven, interpretable, and easily adaptable.
Moreover, unlike previous work, our approach enables rule inference to seamlessly integrate into the library workflow, removing the need to wait for client code migrations. 
We evaluated \tool on pull requests from four popular libraries, successfully mining 461 migration rules from code examples in pull requests and 114 rules from auto-generated code examples. Our generalization procedure increases the number of matches for mined rules by 9$\times$. We applied these rules to client projects and ran their tests, which led to an overall decrease in the number of warnings and fixing some test cases demonstrating \tool's effectiveness in real-world scenarios.

\end{abstract}

\section{Introduction}

Developers often make use of third-party libraries~\cite{de2004good}, which
provide modular functionality to clients through an Application Programming Interface (API).
The API is a contract between the library and its clients, separating the concrete implementation of library features from its specification. 

Ideally, APIs should remain stable. However, they change frequently, and API contracts are often broken \cite{DBLP:conf/wcre/XavierBHV17, DBLP:conf/sigsoft/BogartKHT16, DBLP:journals/smr/DigJ06}, for reasons ranging from bug fixes, to
changes in library requirements~\cite{chapin2001types}. 
APIs may become deprecated or obsolete \cite{perkins2005automatically}, requiring clients to adapt their code to reflect the newest library version. 
These kinds of non-functional code changes are known as \textit{software refactoring} \cite{DBLP:conf/xpu/Fowler02}, a primarily manual \cite{DBLP:conf/sigsoft/KimZN12} and error-prone \cite{DBLP:conf/icse/Kim0DB18} task. 
To migrate to a new library version, clients must examine the library changes such as by inspecting documentation or source code. This task's complexity often deters library clients from updating altogether, despite the security risks posed by outdated dependencies \cite{DBLP:journals/tosem/DilharaKD21} \cite{DBLP:conf/esem/PashchenkoPPSM18}.

The widespread prevalence of deprecations and breaking changes in the software ecosystem motivates research efforts in automating migration~\cite{meditor, soar,apifix,apimigrator,a3,appevolve}. Tools for API migration typically either mine commits from library client projects that have undergone migrations~\cite{apimigrator,a3,meditor}, or are supplemented by information from new client projects in the most up-to-date APIs~\cite{apifix}. 
The effectiveness of these tools is hindered by their reliance on mining data from client projects that have either already migrated across versions, or are already using up-to-date APIs.  Unfortunately, this data is scarce: a recent study found that 81.5\% \cite{developers_dont_update} of projects keep outdated dependencies. Additionally, the mining process can only occur after clients begin transitioning between versions, precluding use shortly after a new version of the library is released \cite{mcdonnell2013empirical, api_evolution_review}.

To overcome these limitations, we propose a new approach called \tool. Unlike previous methods, \tool does not require external data from clients. Instead, it leverages the fact that the development process of open-source libraries provides a wealth of high-quality information that is sufficient to generate transition examples and mine transformation rules. 
At a high level, \textit{our idea is to use pull requests (PRs) submitted to a library's repository to learn code transformation rules for updating client code}. This allows the integration of transformation rule mining into the development process.

Pull requests have become the \textit{de facto} standard for open-source software development on collaborative platforms like GitHub \cite{prs_are_popular, prs_dominate_github}.  
Pull requests typically include a title, a natural language description of the proposed changes and how they relate to project milestones or issues, and a set of commits (i.e., code file changes). These are reviewed by a core group of maintainers who determine to accept, request revisions, or reject the changes. 

We use information from pull requests merged into open-source libraries to mine transformation rules that adapt client code in light of breaking changes or deprecations. First, \tool uses natural language descriptions from PRs to identify API changes by searching for keywords such as ``deprecated", ``breaking change", and ``API change".
If the PR corresponds to such an API change, \tool first \textit{uses the commits in the PRs that contain changes to the library code to generate transformation rules}. The internal updates to the library source code and test cases serve as the ground truth for mining transformation rules.

However, the code-level changes alone do not always provide sufficient information to mine thorough transformation rules for a given breaking change. 
\tool therefore additionally \emph{leverages the natural language text in pull requests to generate additional code examples for mining.}
Specifically, we prompt a state-of-the-art large language model (LLM) pre-trained on open-source code to generate concise code examples that clearly illustrate how to transition from the old API to the new one. Using its prior knowledge of the library (obtained from pretraining) and the additional information in the prompt,   the model can often infer how to transition from old APIs and provide useful concrete examples. 

Using the code examples mined from the library source and the code examples automatically generated from the natural language descriptions, we infer transformation rules in \comby \cite{comby_pldi, comby_website}, a tool and a language for structural code search and replace. We choose to represent our transformation rules in \comby because: (1) it allows us to express find-and-replace rules in a concise and interpretable format; (2) it is a stable widely adopted tool for syntax-driven code transformations. 

There are multiple key advantages to our approach. Firstly, \tool does not require client projects to infer transformation rules.
This contrasts with previous work \cite {DBLP:conf/iwpc/XuDM19, DBLP:conf/icse/FazziniXO20, DBLP:journals/tse/LamotheSC22, DBLP:conf/icse/KetkarSTDB22}, which require large datasets of training data containing multiple migration examples to mine rules. Indeed, \tool can mine migration rules even for changes that have not been in a library release yet (i.e., they are due to future milestones). Secondly, \tool uses \comby to express migration rules, which results in easy-to-interpret, adaptable, and maintainable transformations.

In summary, our main contributions are as follows:

\begin{itemize}
  \item We introduce a novel approach to extract rules that address deprecations and breaking changes in open-source software libraries that does not rely on client data.
  \item An LLM-driven approach for generating code examples and test cases for transformation rule mining.
  \item A generalization procedure for transformation rules enhances their applicability in client projects.
  \item To facilitate integration into existing workflows, we prototype a continuous integration (CI) solution using GitHub Actions for library maintainers, so they can integrate \tool in their workflows.\footnote{\url{https://github.com/squaresLab/melt_action}}
  \item We evaluate \tool on four open-source libraries to infer a total of 461 migration rules from code examples and 114 from auto generated code examples. We also evaluate \tool end-to-end by migrating client code. 
\end{itemize}

\section{Motivation and Overview}
\label{sec:mot}

Figure~\ref{fig:overview} provides a high-level overview of \tool and its main components. We delve into the specifics of each component in Sections~\ref{sec:mining} and \ref{sec:infer}.

\begin{figure}[t]
    \centering
    \includegraphics[width=\linewidth]{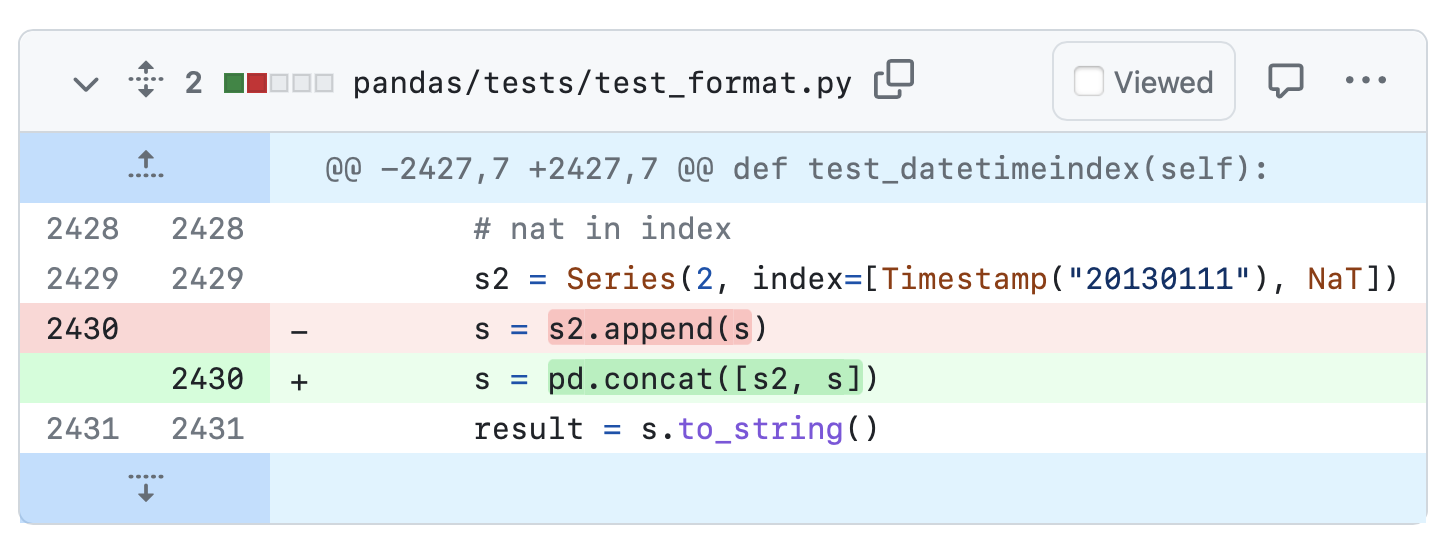}
    \caption{\small Code change in pull request \#44539~\cite{pandas-pr-44539} from the \texttt{pandas-dev/pandas} repository.}
    \label{fig:pr-44539-ex}
\end{figure}

\begin{table}[t]
    \renewcommand{\ttfamily}{\oldttfamily\footnotesize}

    \centering
    \caption{\small \emph{Top:} \comby rules extracted from \texttt{pandas} pull request  \#44539, deprecating \texttt{DataFrame.append} and \texttt{Series.append}. \emph{Bottom:} Rules extracted from \texttt{sci-py} pull request \#14419,
    including original specific (``Spec'') and generalized (``Gen'') versions. Template variable constraints are omitted for brevity.}
    \label{tab:comby-rules-append}
    \begin{tabular}{>{\raggedright\arraybackslash}p{0.8cm} >{\raggedright\arraybackslash}p{3.6cm} >{\raggedright\arraybackslash}p{3cm}}
    \toprule
   &  \textbf{Match Template} & \textbf{Rewrite Template} \\
    \midrule
  &  \multirow{2}{*}{\lstinline|*:[[s2]]*.append(*:[[s1]]*)|} & \lstinline|pd.concat([*:[[s2]]*, *:[[s1]]]*)| \\
   & \multicolumn{1}{r}{\lstinline|where *:[[s1]]*.type == Series|} &\\
   & \multicolumn{1}{r}{\lstinline|*:[[s2]]*.type == Series |} &\\
    \midrule
   & \multirow{3}{*}{\lstinline|*:[[df]]*.append(*:[[s]]*)|} & \lstinline|pd.concat([*:[[df]]*, DataFrame(*:[[s]]*).T.| \\
   &  & \lstinline|infer_objects()])| \\
   & \multicolumn{1}{r}{\lstinline|where *:[[df]]*.type == DataFrame|} &  \\
   & \multicolumn{1}{r}{\lstinline|*:[[s]]*.type == Series|} &  \\
    \bottomrule
        \toprule
    \textbf{Type} & \textbf{Match Template} & \textbf{Rewrite Template} \\
    \midrule
    \multirow{2}{*}{Spec} & \lstinline|*:[[s]]*.spline.| & \lstinline|*:[[s]]*.cspline2d(| \\
    & \lstinline|cspline2d(*:[[x]]*,*:[y])*| & \lstinline|*:[[x]]*, *:[y]*)| \\
    \midrule
    \multirow{2}{*}{Gen} & \lstinline|*:[[s]]*.spline| & \lstinline|*:[[s]]*.cspline2d(| \\
    & {\lstinline|cspline2d(*:[args]*)|} & \lstinline|*:[args]*)| \\
    \bottomrule
    \end{tabular}
\end{table}

\renewcommand{\ttfamily}{\oldttfamily\small}

\begin{figure*}[t]
  \centering
  \includegraphics[width=\linewidth]{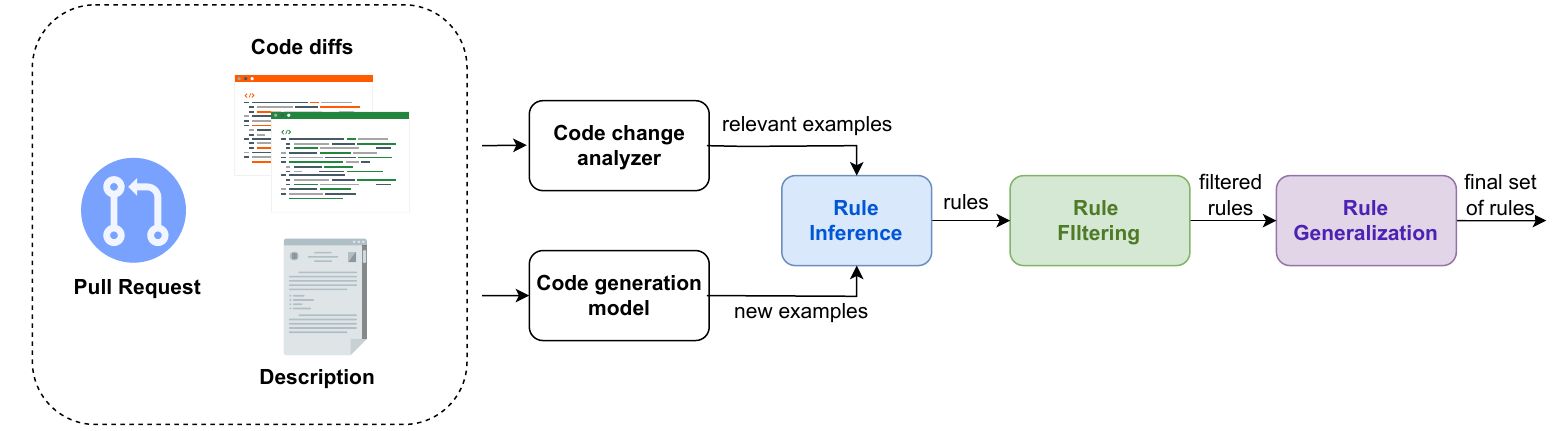}
  \caption{\small \tool overview. \tool takes as input a pull request (PR) and outputs a set of rules. The PR is processed in two ways: (1) the Code change analyzer identifies relevant code changes; (2) the Code generation model generates additional code examples. Rules are inferred from the code changes and examples using the rule inference algorithm, then filtered and generalized.}
  \label{fig:overview}
\end{figure*}

Pull requests are the input of \tool, as they are the key source that informs our approach.
Pull requests generally contain all the code changes related to a given new feature.
For example, Figure~\ref{fig:pr-44539-ex} shows an example code change from a pull request~\cite{pandas-pr-44539} submitted to \pandas~\cite{pandas-github} that deprecates two popular APIs: \texttt{DataFrame.append} and \texttt{Series.append}.\footnote{Both APIs were later removed from \pandas in version $2.0.0$.}  
\tool identifies code changes, such as the one shown in Figure~\ref{fig:pr-44539-ex}, within the pull request using its \textit{Code Change Analyzer} (Section \ref{sec:miningdiff}) and inputs them into the \textit{Rule Inference} algorithm (Section \ref{sec:inferc}) to generate rules. The top portion of Table \ref{tab:comby-rules-append} shows two of the rules \tool infers from the code changes for this specific pull request. 

The rules in Table \ref{tab:comby-rules-append} are expressed in \comby's domain specific language \cite{comby_website}. 
The match template (left column) is the code structure for which \comby searches. 
The rewrite template (right column) shows how to transform the matched code based on the variables in the match template \cite{comby_pldi}. 
\comby uses template variables, i.e., placeholders that can be matched with certain language constructs. For example, a template variable to match alphanumeric characters is represented by \lstinline|*:[[x]]*|, where \lstinline|*x*| is the name of the template variable. 
The template variables in the match template can be constrained in multiple ways using a \texttt{where} clause. 
In particular, to prevent spurious matches, template variables can be constrained to be a certain type (like \lstinline|*:[[s2]]*.type == DataFrame|).
Although type information is not strictly required, it is especially useful when working with common API names such as \texttt{append} and \texttt{concat}, since both are part of Python's builtins list.

Code diffs in pull requests provide valuable information, however, they do not always contain the necessary code examples for rule inference.
Fortunately, pull requests offer alternative sources of information that can be used to extract further details about the changing APIs. 
Figure \ref{fig:spline} shows an informative comment left by a developer in a code file when deprecating namespace \scipy's~\cite{scipy} namespace \lstinline|scipy.signal.spline| in favor of \lstinline|scipy.signal|.
To leverage all available information in the pull request, \tool uses a \textit{Code Generation Model} to generate additional code examples and test cases for this change (Section~\ref{sec:mininggen}).
Figure~\ref{fig:code_example} shows a simplified version of code \gpt \cite{openai-gpt4} (a state-of-the-art model) generates from the pull request in Figure~\ref{fig:spline}. The generated examples enable us to both infer and test the rules.

\begin{figure}[t]
    \centering
    \includegraphics[width=\linewidth]{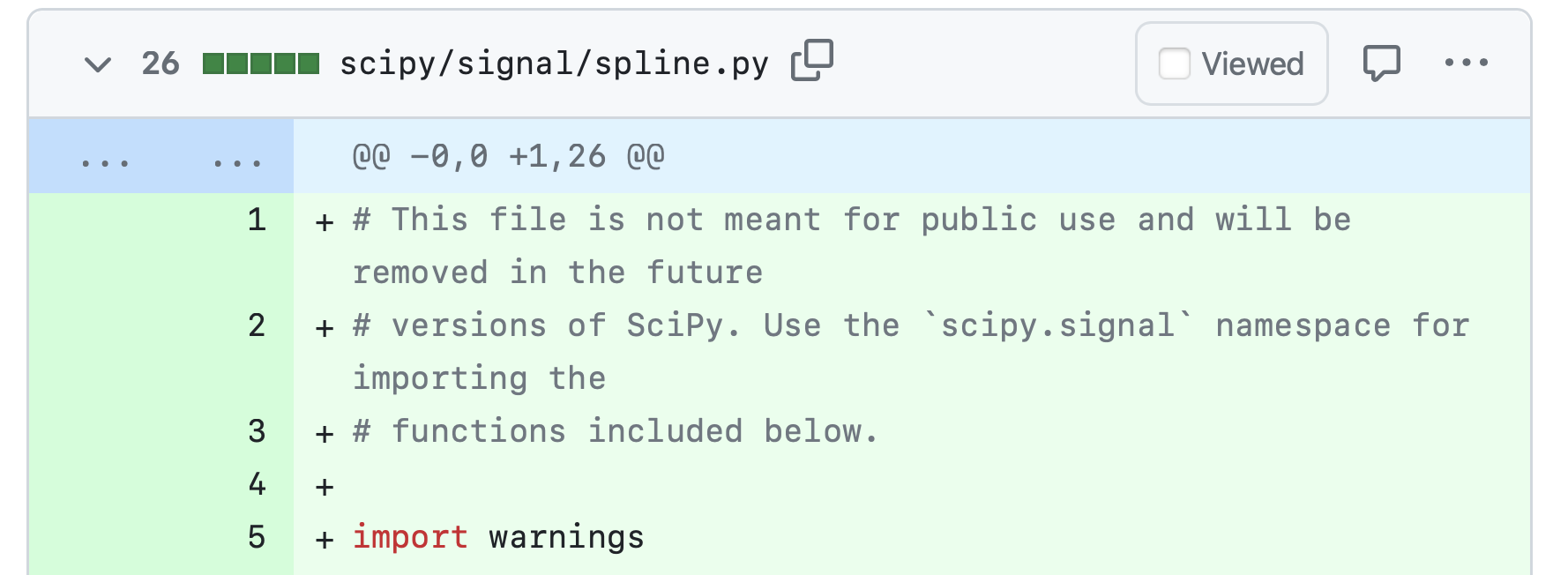}
    \caption{\small Pull Request \#14419 \cite{scipy-github-pr-14419} from \texttt{scipy/scipy}. This pull request was part of SciPy 1.8, released in February 2022.}
    \label{fig:spline}
\end{figure}

\begin{figure}[t]
    \centering
    \begin{lstlisting}[style=pythonstyle]
def old_usage1(image):
  return signal.spline.cspline2d(image, 8.0)

def new_usage1(image):
  return signal.cspline2d(image, 8.0)

class TestEquiv(unittest.TestCase):
  def test_assert1(self):
    np.random.seed(181819142)
    image = np.random.rand(71, 73)
    assert np.allclose(
            old_usage1(image), 
            new_usage1(image))
    \end{lstlisting}
    \caption{\small Code generated by \gpt showing how to transition from the deprecated namespace for \lstinline|cspline2d| and a test case.}
    \label{fig:code_example}
\end{figure}

Since the test case executes successfully, \tool uses the code example to generate a rule by abstracting concrete identifiers and literals. 
For this case, \tool generates the rule in the third row of Table \ref{tab:comby-rules-append}. 
This rule accurately reflects the deprecation made in the pull request (i.e., replaces the deprecated namespace with the new one). 
Nevertheless, a closer inspection reveals that the rule is too specific:
it will only match usages where: (1) the first argument of \texttt{cspline2d} is an identifier (\lstinline|*:[[s]]*| only matches with identifiers), and (2) the function is called with two or more arguments. 
The \texttt{cspline2d} function can accept multiple combinations of arguments, including keyword arguments with default values.

To guard against overly-specific rules, \tool applies \textit{Rule Generalization}  (Section~\ref{sec:gen}).
For example, the template holes \lstinline|*:[[x]]*| and \lstinline|*:[y]*| in the rule in the first row of the bottom of Table \ref{tab:comby-rules-append} remain unchanged in the match and rewrite templates, indicating that they are not relevant to the change at hand. 
To enhance the rule's applicability, \tool generalizes the specific argument combination, resulting in an updated version of the rule (shown in the last row of Table \ref{tab:comby-rules-append}). The revised rule uses a more permissive match template using \lstinline|*:[args]*|, which can match any number of function arguments.

\section{Mining Pull Requests}
\label{sec:mining}

In this section, we describe \tool's approach to identify and create code examples for rule inference.

\subsection{Extracting Code Examples from Code Diffs}
\label{sec:miningdiff}

\tool's input is a pull request $\mathcal{P}$, which contains both natural language and a set of code changes / diffs $\mathcal{P}.\textit{diffs}$, each corresponding to changed code snippets. However, not all diffs in a PR are relevant to an API change, as they may encompass unrelated refactoring actions. Therefore, \tool first identifies which diffs in the PR are relevant to the API of interest.

\tool determines which code changes are relevant using its \textit{Code Change Analyzer}. This works in two steps: (1) identifying relevant public API names affected in the PR, and (2) filtering the code diffs that contain example usages of such APIs. First, to identify the public APIs affected in the PR, \tool analyzes each diff to determine its enclosing function and class,  thereby compiling a list of affected functions. Then, \tool filters out non-relevant API names by excluding test functions and private namespaces from this list. For example, for the code change in Figure \ref{fig:pr-44539-ex}, \tool does not classify the function name \texttt{test\_datetimeindex} as an API of interest because it is test-related. However, other changes in the same PR (not shown) affect the \texttt{append} and \texttt{concat} public methods, so \tool considers those as APIs of interest. This step yields a list of function names that are considered APIs of interest.

Next, \tool filters the code diffs using the list of APIs identified in the previous step. It retains only those diffs that contain usages of these APIs. For example, the code change in Figure \ref{fig:pr-44539-ex} is kept because it \emph{does} contain usages of \texttt{append} and \texttt{concat}, even though the test method itself is not a relevant API name. A strength of this approach is its generalizability across multiple libraries and languages. This step results in a filtered set of code diffs considered relevant  for rule inference.

\subsection{Generating Additional Code Examples}
\label{sec:mininggen}

As illustrated in Section \ref{sec:mot}, pull requests sometimes lack sufficient code examples to infer migration rules. 
In a preliminary study, we analyzed 174 PRs related to breaking changes and deprecations from \pandas' release notes. 
We discovered that only $41$ ($23.6\%$) of these PRs contained at least one meaningful code example showcasing the transition from old to new usage.
However, PRs offer other information sources about API changes, including  natural language in comments, developer discussions, and documentation.
Our key insight is that this additional data can also be leveraged to generate and test more code examples. 
\tool uses a \textit{Code Generation Model} to produce extra code examples from this data. 
Generating code examples rather than rules directly is advantageous because the generated code can be tested and validated, enhancing confidence in the inferred rules.
Additionally, the code examples may enhance interpretability by demonstrating the provenance of inferred rules to users. 

Algorithm \ref{alg:gen_gpt} outlines our approach. 
Given a PR and a code generation model, \tool iterates for a fixed number of times $N$ (based on the desired number of samples) and asks the model to generate a transition example (line 3).
Our \textsc{generateExample} implementation prompts \gpt~\texttt{8K}~\cite{openai-gpt4}, which is well-versed in our target libraries' code, to process PR information (code diffs, title, description, discussion) and generate transition examples for the APIs affected in that PR.
Figure \ref{fig:prompt1} shows the template used for the prompt.\footnote{Full prompts are provided in the artifact at Zenodo~\cite{melt_artifact}.}
\tool uses the model to generate a pair of code examples, $e_{old}$ and $e_{new}$, representing the old and new usages, respectively. While $e_{old}$ uses the old API, $e_{new}$ is implemented using the new API.
Both examples are functions with identical signatures but different implementations. 

\begin{figure}[t]
    \centering
    \includegraphics[width=\linewidth]{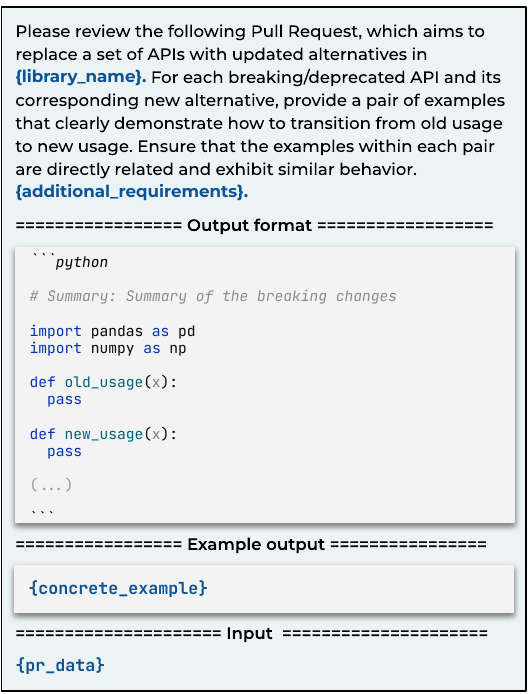}
    \caption{\small Prompt template for the \textsc{generateExample} function in Algorithm 1, featuring four placeholders: (1) \texttt{library\_name}, (2) \texttt{additional\_requirements} for format consistency and correctness, (3) a \texttt{concrete\_example} with summary and examples from \pandas, and (4) \texttt{pr\_data}, the PR information including title, description, changed files, and corresponding diffs, as JSON.}
    \vspace{-0.5cm}
    \label{fig:prompt1}
\end{figure}

\begin{figure}[t]
    \centering
    \includegraphics[width=\linewidth]{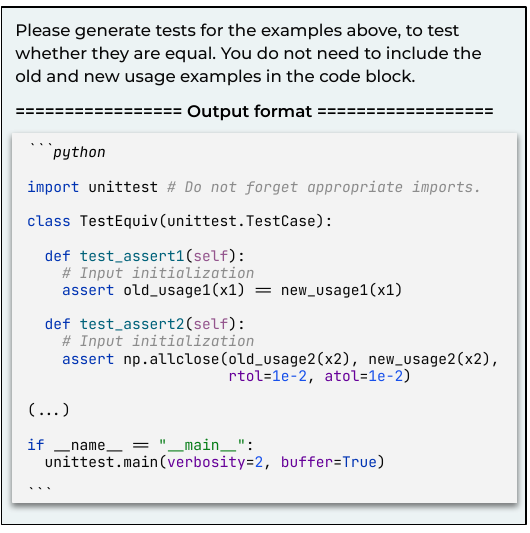}
    \caption{\small Test case generation prompt. \tool concatenates the prompt from Figure \ref{fig:prompt1}, the model's response, and this prompt to ask the model for tests for the generated examples.}
    \label{fig:prompt2}
\end{figure}

\begin{algorithm}[t]
    \caption{\textsc{GenerateTransitionExamples}($P, \mathcal{M}, N$)}
    \label{alg:gen_gpt}
    \begin{algorithmic}[1]
    \Require $\mathcal{P}$: PR, $\mathcal{M}$: gen model, $N$: number of samples
    \Ensure $E$: transition examples
    
    \State {$E \gets \emptyset$}
    
    \For{$i = 1$ \textbf{to} $N$}
        \State {$(e_{old}, e_{new}) \gets \textsc{generateExample}(\mathcal{M}, \mathcal{P})$}
        \State {$\mathcal{T}_E \gets \textsc{generateTestCases}(\mathcal{M}, \mathcal{P}, e_{old}, e_{new})$}
        \State {$E \gets E \cup \{(e_{old}, e_{new})\}$}
        \ForEach{$\text{test} \in \mathcal{T}_E$}
            \If{$\textsc{fails}\text{(test)}$}
                \State {$E \gets E \setminus \{(e_{old}, e_{new})\}
                $}
            \EndIf
        \EndFor
    \EndFor
    
    \State \Return $E$
    
    \end{algorithmic}
\end{algorithm}

However, simply asking the model to generate a code example is not enough, as there are no guarantees that $e_{new}$ has the same semantics of $e_{old}$. 
As a subsequent step, \tool generates test cases that assess the equivalence between $e_{old}$ and $e_{new}$ (line 4).
In our implementation of \textsc{generateTestCases}, \tool follows up with \gpt for test generation. The request includes the original prompt, the model's response, and the text from Figure \ref{fig:prompt2}. \gpt generates test inputs and computes their output on $e_{old}$, which serve as an oracle to test $e_{new}$.
The test case asserts that $e_{new}$ produces the same output as $e_{old}$ for the same set of inputs.

After generating test cases, \tool checks whether any test fails (lines 6-10). If any test case fails, the transition examples for that test are discarded (line 8), as the new usage does not behave similarly to the old one. \tool only considers examples for which test cases were generated. This procedure outputs a set of transition examples (when possible) that can then be used to infer migration and test rules.

\section{Rule Generation}
\label{sec:infer}

\tool uses the \comby language~\cite{comby_pldi} and toolset for searching and refactoring source code~\cite{comby_website} to express migration rules. We introduced some elements of the language in Section~\ref{sec:mot}, with examples of \comby's syntax-driven match and rewrite templates.
Formally, a rewrite rule in \comby is of the form \lstinline[mathescape]|M $\xrightarrow{}$ R where c$_1$, c$_2$, ..., c$_n$|, where \lstinline|M| is the match template, \lstinline|R| is the rewrite template, and c$_1$, c$_2$, ..., c$_n$ are constraints in the rule language. 
The key structure of \comby rules are template variables, which are 
holes in the match and rewrite templates that can be filled with code. 
Template variable types include, e.g., \lstinline|*:[[x]]*| matching alphanumeric characters (similar to \lstinline|\w+| in regex), and \lstinline|*:[x]*| matching anything between delimiters (e.g., \lstinline|[],(),{}|).
\comby also supports a small rule language to add additional constraints, like types or regular expression matches, on the template variables.
\comby's website~\cite{comby_website} provides the full syntax reference.
Although language agnostic, \comby is still language \emph{aware}, and can deal with comments and other language-specific constructs.  
Its rules are also close to the underlying source, and thus typically easier to read than, e.g., transformations over abstract syntax trees. 

The rest of this section describes rule inference. 

\subsection{Rule Inference}
\label{sec:inferc}

Given a set of code examples, \tool infers a set of \comby rules that can be used to automatically migrate APIs in client code.
First, \tool parses the code files corresponding to each code diff into an abstract syntax tree (AST), identifying the nodes corresponding to the change before and after. 
\tool then uses a variation of \inferrule's algorithm \cite{inferrule} (adapted to Python) that always returns a single rule, and never abstracts away class names, method names, and keyword arguments.

\begin{figure}[t]
  \centering
    \begin{subfigure}[t]{0.22\textwidth}
      \caption{Code before migration}
      \begin{lstlisting}[style=pythonstyle,
        stepnumber=1, mathescape, escapechar=$]
r = pd.read_csv(
  filename,
  compression=comp,
  encoding=enc,
  index_col=0,
$\colorbox{red!30}{- squeeze=True)}$
      \end{lstlisting}
    \end{subfigure}
    \hfill
    \begin{subfigure}[t]{0.22\textwidth}
      \caption{Code after migration}
      \begin{lstlisting}[style=pythonstyle2,
        stepnumber=1, mathescape, escapechar=$]
r = pd.read_csv(
  filename,
  compression=comp,
  encoding=enc,
  index_col=0).
$\colorbox{green!30}{+ squeeze()}$
      \end{lstlisting}
    \end{subfigure}
    \caption{\small Example code change from PR \#43242 \cite{pandas-pr-43242} in \pandas}
    \label{fig:squeeze}
    \vspace{-0.5cm}
  \end{figure}


To illustrate, consider the code change in Figure \ref{fig:squeeze}, where a library maintainer transforms a keyword argument into a function call. 
The smallest unit \tool considers for a \comby rule is a source code line. Given the two assignment nodes corresponding to the change, rule inference then abstracts away child nodes with template variables.
When a construct has the same character representation, \tool uses the same template variable. For the example, \tool abstracts the left-hand side and right-hand side of both assignments, yielding:

\lstinline[mathescape]|*:[[a]]*  = *:[b]*|, and \lstinline[mathescape]|*:[[a]]*  = *:[c]|.

Notice that the template variable for the target of both assignments is the same, \lstinline[mathescape]|*:[[a]]*|, because their source representation is the same.
However, \tool cannot match the right-hand side of the assignments (\lstinline|*:[[b]]*|, and \lstinline|*:[[c]]*|).  It, therefore further decomposes the AST nodes'  children:

\begin{minipage}[t]{0.26\textwidth}
    \begin{lstlisting}[style=pythonstyle, mathescape, escapechar=$]
*:[[a]]* = 
  *:[[i]]*.read_csv(
  *[[d]]*,
  compression=*:[e]*,
  encoding=*:[f]*,      $\xrightarrow{}$
  index_col=*:[[g]]*,
  squeeze=*:[[h]]*)
    \end{lstlisting}
    \end{minipage}
    \hfill
    \begin{minipage}[t]{0.20\textwidth}
    \begin{lstlisting}[style=pythonstyle2, mathescape, escapechar=$]
*:[[a]]* = 
  *:[[i]]*.read_csv(
  *:[[d]]*,
  compression=*:[e]*,
  encoding=*:[f]*,
  index_col=*:[[g]]*)
 squeeze()
    \end{lstlisting}
\end{minipage}

Notice that, as previously discussed, \tool never abstracts away class names, function names, and keyword arguments, as preserving these details is crucial for API migration.
In this case, \tool can match every template variable in the match template with a corresponding variable in the rewrite template except \lstinline|:*[[h]]*|. Consequently, it attempts to further decompose the nodes, but still fails to match \lstinline|*:[[h]]*|, ultimately reverting it and generating the final rule:

\begin{minipage}[t]{0.26\textwidth}
  \begin{lstlisting}[style=pythonstyle, mathescape, escapechar=$]
*:[[a]]* = 
  *:[[i]]*.read_csv(
  *[[d]]*,
  compression=*:[e]*,
  encoding=*:[f]*,      $\xrightarrow{}$
  index_col=*:[[g]]*,
  squeeze=True)

where
  *:[[i]]*.type == pandas
  \end{lstlisting}
  \end{minipage}
  \hfill
  \begin{minipage}[t]{0.20\textwidth}
  \begin{lstlisting}[style=pythonstyle2, mathescape, escapechar=$]
*:[[a]]* = 
  *:[[i]]*.read_csv(
  *:[[d]]*,
  compression=*:[e]*,
  encoding=*:[f]*,
  index_col=*:[[g]]*)
 squeeze()
  \end{lstlisting}
\end{minipage}

After inferring a rule, \tool incorporates type guards. 
The goal is to constrain each template hole to its respective observed type. 
This step is crucial in preventing the misapplication of rules for common API names (e.g., matching \lstinline|List.append| when the rule targets \lstinline|DataFrame.append|). 
In contrast to previous rule synthesis approaches \cite{inferrule,pyevolve}, \tool directly incorporates type constraints into \comby's rule language.
This integration is possible because we extend \comby to support Language Server Protocol (LSP) type inference.
\tool uses the \jedi \cite{jedi} type inference language server, making it available for client usage.

\subsection{Rule Filtering}
\label{sec:filtering}

\tool sometimes generates \emph{duplicate rules}. A rule is considered a duplicate if the match template, rewrite template, and template variable constraints are identical. Therefore, \tool first discards \emph{duplicate rules} within the same PR (post-generalization as well). Occasionally, \tool also infers irrelevant and spurious rules (e.g., rules that contain variables in the rewrite template that might not be in scope). Thus, to return high-quality rules, \tool filters by:

\subsubsection{API Keywords} \tool discards transformation rules that do not contain the name of any affected APIs. This can occur when a developer modifies the surrounding context of a code block, for example, by wrapping a statement in a try-catch block (e.g., \lstinline[mathescape]|*:[x]* $\xrightarrow{}$ try:\n*:[x]*|). These rules are considered spurious because they can match arbitrary code and are not specific to API migration. 

\subsubsection{Unsafe Variable and Private Namespaces} \tool discards rules where a rewrite template uses either variables from private namespaces (indicated by calls with underscores, Python's convention for private attributes/functions/namespaces), or variables not present in the match template. 
This ensures that the rules do not rely on private or internal functionality that is not accessible to client code.

\subsection{Generalizing Rules}

\label{sec:gen}

Rules inferred from single code examples can often be too narrow in scope, as seen in the \texttt{squeeze} example. This rule uses a specific set of arguments for \lstinline|read_csv|. However, since \lstinline|read_csv| includes many optional arguments, the essence of the change should focus on altering the \lstinline|squeeze| parameter, regardless of other arguments. Additionally, the current rule only applies to assignments, even though it concerns API usages that may appear in various contexts, not just assignments.

Therefore, our approach generalizes rules for broader applicability by abstracting irrelevant context and generalizing arguments. 
Algorithm \ref{alg:rw} overviews the process.
\tool obtains AST nodes corresponding to the match and rewrite templates (lines 1-2), and isolates and eliminates all constructs unrelated to the actual code transformation (line 3). Specifically, \textsc{removeCommonContext} unwraps return statements, removes targets on assignments (when possible), and unwraps conditionals, asserts, and other statements, provided they are identical in both the match and rewrite templates. If there are multiple ways to unwrap a statement (e.g., the rule is comprised of two assignments statements), \tool returns the first possible unwrapping.

Next, API call arguments are generalized wherever possible. \tool uses matchings obtained from the Hungarian algorithm during the rule inference process (further explained in \cite{inferrule}) to find matchings between call nodes. \tool examines matching call nodes and generalizes common arguments (line 5).
The \textsc{generalizeArguments} function operates by examining pairs of arguments and keyword arguments. 
If there are multiple consecutive arguments between the match and rewrite templates, we replace the arguments with a generic template variable \lstinline|*:[args]*|. Once the arguments of the call pair have been generalized, \tool replaces it in the original templates (lines 6-7).
\tool also ensures that keyword arguments always appear at the end of the rewrite template.
For example, when a developer turns a positional argument into a keyword argument, the rewrite template moves the positional argument to the position of the last keyword argument. For our running example, the final rule is:

\begin{minipage}[t]{0.28\textwidth}
    \begin{lstlisting}[style=pythonstyle, mathescape, escapechar=$]
*:[[i]]*.read_csv(
  *:[args]*,         $\xrightarrow{}$
  squeeze=True)

where *:[[i]]*.type == pandas
    \end{lstlisting}
    \end{minipage}
    \hspace{-0.8cm}
    \begin{minipage}[t]{0.17\textwidth}
    \begin{lstlisting}[style=pythonstyle2, mathescape, escapechar=$]
*:[[i]]*.read_csv(
  *:[args]*)
 squeeze()
    \end{lstlisting}
\end{minipage}

\noindent \ldots where \tool removed the assignment target and abstracted irrelevant arguments. 

Generalization is crucial to ensuring broader rule applicability. 
However, over-generalization does occur, especially when type information is lost. 
As a result, generalized rules may need extra validation.
\tool does allow users to generate rule variations to explore alternative generalizations. For example, a rule with a match template \lstinline[style=pythonstyle]|*:[[i]]*.read_csv(squeeze=True)|, could have a variation with more arguments before / after \texttt{squeeze}, i.e., \lstinline|*:[[i]]*.read_csv(*:[args0]*, squeeze=True, *:[args1]*)|. 
We leave a detailed investigation of these concerns to future work.

\begin{algorithm}[t]
    \caption{\textsc{Generalize}(r)}
    \label{alg:rw}
    \begin{algorithmic}[1]
    \Require $r$: a rewrite rule 
    \Ensure generalized rewrite rule

    \State {$n_1 \gets \textsc{getBeforeNode}(r)$}
    \State {$n_2 \gets \textsc{getAfterNode}(r)$}
    
    \State {$n_1, n_2 \gets \textsc{removeCommonContext}(n_1, n_2)$}



    \For{$(c_1, c_2) \in \textsc{getCallPairs}(n_1, n_2) $}
        \State {${c_1}^\prime, {c_2}^\prime  \gets \textsc{generalizeArguments}(c_1, c_2)$}
        \State {$n_1 \gets \textsc{replaceNode}(n_1, c_1, c_1^\prime)$}
        \State {$n_2 \gets \textsc{replaceNode}(n_2, c_2, c_2^\prime)$}
    \EndFor

    \State \Return $\textsc{CreateRule}(n_1, n_2)$
    
    \end{algorithmic}
\end{algorithm}

\section{Evaluation}

We answer the following research questions:

\begin{enumerate}[label=\textbf{{RQ}\arabic{*}.}, leftmargin=*,widest=10]
  \item How effectively can \tool generate transformation rules from code examples in PRs?
  \item How do code examples generated automatically complement code examples in PRs?
  \item What is the impact of rule generalizability?
  \item Are the rules effective for updating client code?
\end{enumerate}

\subsection{Experimental Setup}

\subsubsection{Implementation}

Although our approach is largely language-agnostic, we implement it for Python libraries because: (1) Python is one of the most popular programming languages~\cite{tiobe}, and (2) there exists a gap in migration tools for Python~\cite{api_evolution_review}.
We implemented rule inference using the Python abstract syntax tree (AST) module.
\inferrule \cite{inferrule} was originally implemented for Java AST; we brought native implementation to Python.
We also perform rule generalization at the Python AST level.
For code generation, we used \mbox{\gpt}~\cite{openai-gpt4} (version \texttt{gpt-4-0314}).
We extended \comby to support Language Server Protocol (LSP)-based type inference over match templates~\cite{comby-types} with \jedi \cite{jedi}, a state-of-the-art static analysis tool. \tool's source code, data, and logs used for the evaluation are available at Zenodo \cite{melt_artifact}.

\subsubsection{Methodology}
\label{sec:evalmethodology}

We evaluated \tool using four of the most popular Python data science libraries: \numpy, \scipy, \sklearn, and \pandas.
We collected a total of 722 PRs for \pandas, 141 for \sklearn, 186 for \numpy, and 130 for \scipy using the \texttt{GitHub} API and web crawlers over release notes. 
We took a convenience sampling approach to find PRs concerning API breaking changes, or deprecation-related PRs, moving backwards from the version of each library (as of April 2023); this includes merged PRs intended for future library releases, as well as those that have been released. 
We collected more PRs for \pandas than other libraries because it had a higher number of PRs, and breaking changes in \pandas are particularly well documented. We then executed \tool on each PR. 

For our manual assessment of rule correctness and relevancy, two authors of this paper manually labeled a set of rules independently.
We define a \emph{correct rule} as one that (1) accurately reflects the changes in the PR and (2) is generally applicable to client code without overgeneralizing (i.e., it will not produce incorrect migrations even if the rule works correctly in some cases).
This process requires analyzing the PR discussion, changes, source code, and documentation when necessary. 
The annotators discussed five representative examples together and then individually labeled 151 unique rules, achieving an inter-rater reliability (IRR) with a Cohen's kappa of $0.84$ (almost perfect agreement)~\cite{cohen_coefficient_1960}.
Due to the high agreement, the first author labeled the remaining rules to cover all research questions.

\subsection{RQ1: Mining Rules from Code Examples in PRs}
\label{sec:exeval}

\begin{table}[t]
  \caption{\small \textbf{RQ1.} \emph{Left:} PRs per library, with mined rules and correct rules. \emph{Right}: Filtered and generalized rules mined per library, with total and correct counts.}
  \label{tab:mining_rules}
  \centering
  \small
  \begin{tabular}{lrrr|rr}
  \toprule
  & & \multicolumn{2}{c}{\textbf{PRs with}} & \\
  \multirow{2}{*}{\textbf{Library}} & \multirow{2}{*}{\textbf{\# PRs}} & \textbf{Mined} & \textbf{Correct} & \multicolumn{2}{c}{\textbf{Mined Rules}} \\
  & & \textbf{Rules} & \textbf{Rules}  &  \textbf{Total} & \textbf{Correct (\%)} \\ \midrule    
  \pandas & 722 & 169 & 102 & 521  & 359 (\SI{68.9}{\percent}) \\
  \scipy & 130 & 21 & 11 & 33  & 19 (\SI{57.6}{\percent})\\
  \numpy & 186 & 20 & 10   & 47  & 27 (\SI{57.4}{\percent})\\
  \sklearn & 141 & 38  & 21 & 82  & 56 (\SI{68.3}{\percent})\\
  \midrule
  \textbf{Total} & 1179& 248 & 144 & 683 & 461 (\SI{67.5}{\percent})\\    \bottomrule
  \end{tabular}
\end{table}


Table~\ref{tab:mining_rules} summarizes \tool's rule inference algorithm on 1179 PRs (722 \pandas, 130 \scipy, 186 \numpy, 141 \sklearn). 
\tool's ability to extract code examples from PRs largely depends on the libraries' testing practices. Nonetheless, a significant number of PRs contain valuable examples for rule extraction. Previous studies \cite{impossible_migration} found that only 27.1\% of migrations in a different set of libraries were potentially fully automatable. \tool generates correct migration rules for 12.2\% of analyzed PRs, indicating room for improvement (further explored in RQ2).



Running \tool's rule inference algorithm to the 1179 PRs results in 5504 rules. After filtering and generalization, we ended up with 683 rules.
The right-most columns of Table \ref{tab:mining_rules} show the number of mined rules after generalization and filtering for each library, and their correctness based on manual validation. On 67.5\% of the cases, our mined rules are correct and do not overgeneralize. 
However, on 32.5\% of the cases, \tool derived incorrect, non-generally applicable rules. We observed three primary reasons for incorrect rules: (1) 
\emph{Code change not generally applicable}, such that the rule does not capture the context in which it is applicable. For example, in \numpy PR \#9475 \cite{numpy-pr-9475}, \texttt{np.rollaxis} is replaced in favor of \texttt{np.moveaxis}. The correct way to migrate depends on the specific arguments of \texttt{np.rollaxis}, as its behavior varies based on these arguments. Our rule cannot capture this, as it only constraints variables over types, not their content; (2) \emph{Overgeneralization} of rule arguments. 
 For instance, \pandas PR \#21954 \cite{pandas-pr-21954} states ``\textit{read\_table is deprecated. Instead, use pandas.read\_csv passing sep=`t' if needed.}" However, one of the inferred rules is \lstinline|read_table(*:[args]*)| $\mapsto$ \lstinline|read_csv(*:[args]*)|. Our algorithm abstracts all arguments based on the code example, not accounting for when \texttt{sep=`t'} is necessary;
(3) \emph{Unrelated} rules that had not been filtered correctly.

\begin{highlight}
  \tool generates \textbf{461 correct migrations rules} directly \textbf{from code examples for 144 (12.2\%) out of 1179 PRs} from four popular data-science libraries.
  \end{highlight}

\subsection{RQ2: Automatically generated code examples}
\label{rq:lm}

To assess the impact of example generation on rule inference, we sampled 50 PRs from each library (limited by budget).
We used the templates from Figures \ref{fig:prompt1}, and \ref{fig:prompt2} to prompt the model to generate both code examples and test cases/inputs for each PR. The prompt includes the title, description, discussion, and code changes of the PR.
We prompted \gpt, with a (default) temperature of $0.2$, and sampled the model with $N=5$ in Algorithm \ref{alg:gen_gpt}.

The left side of Table \ref{tab:mining_rules_rq2} shows the number of unique examples generated for each library and the number of examples that passed the automatically generated test suite.
From these examples, \tool inferred 248 unfiltered and ungeneralized rules; filtering and generalization reduced the set to \emph{156 unique rules}. 
We also assessed whether these rules could have been generated from the PR code directly, by checking (1) whether they were mined in RQ1 (Section~\ref{sec:exeval}), or (2) whether they could be directly applied to their corresponding PR (meaning that they \emph{could} have been mined in RQ1, but may have been heuristically filtered away).  

\begin{table}[t]
  \caption{\small \textbf{RQ2.} \emph{Left:} Code examples generated and passing tests per library. \emph{Middle:} PRs with mined and correct (``Corr.'') rules from generated examples. \emph{Right:} Filtered and generalized rules per library. \emph{Note:} Limited to 50 PRs per library for budgetary reasons.}
  \newcommand\mc[2]{\multicolumn{#1}{c}{\textbf{#2}}}
  \label{tab:mining_rules_rq2}
  \centering
  \small
  \resizebox{\columnwidth}{!}{%
  \begin{tabular}{lrr|rr|rrr}
  \toprule
   & \mc{2}{Code}  & \mc{2}{PRs with}   & \mc{3}{Mined Rules}\\
 & \mc{2}{Examples} &\mc{2}{Rules} & &\multicolumn{2}{c}{\textbf{Correct}} \\
  \textbf{Library} & \textbf{Total} & \textbf{\# pass} & \textbf{Total} & \textbf{Corr.}      & \textbf{Total} &  \textbf{Prev} & \textbf{New}   \\ \midrule  
  \pandas & 285 & 134 & 25  & 19  & 45  & 7       & 30 \\
  \scipy & 194 & 68 & 15   & 13  & 30  & 4        & 18 \\
  \numpy & 222 & 114 & 21   & 14 & 46   & 2        & 31  \\
   \sklearn & 187 & 63 & 21 & 13  & 35 & 5        & 17 \\
  \midrule
  \textbf{Total} & 888 & 379 & 82 & 59 & 156 & 18 & 96 \\    \bottomrule
  \end{tabular}}
\end{table}

Table \ref{tab:mining_rules_rq2} summarizes rule mining success using the generated examples per PR (middle columns); the right-hand side shows the number of rules mined.  We categorized the correct rules into two groups: those that could have been mined without the new examples (prev), and those that could only be inferred with the generated examples.
Like in the RQ1, \tool can generate incorrect rules in some scenarios. Consider the following example rule:
\lstinline[mathescape, basicstyle=\ttfamily\fontfamily{zi4}\selectfont]|*:[[y]]*.shift(*:[x]*, fill_value=*:[z]*)|
$\xrightarrow{}$ \lstinline[mathescape, basicstyle=\ttfamily\fontfamily{zi4}\selectfont]|*:[[y]]*.shift(*:[x]*, fill_value=pd.Timestamp(*:[z]*))|~\footnote{The types of the template variables are omitted for brevity.}. This rule is derived from \pandas PR number \#49362 \cite{pandas-pr-49362}. The release notes for the PR state:
\textit{``Enforced disallowing passing an integer fill\_value to DataFrame.shift and Series.shift with datetime64, timedelta64, or period dtypes"}. This transformation is only valid if the \texttt{Series} is of \texttt{datetime64 objects}, a condition not captured by the rule. 
While the transformation correctly preserves behavior for series containing \texttt{datetime64 objects}, it is incorrect for general application.
More diverse tests for the code example could likely increase coverage and filter more incorrect rules. 


\begin{highlight}
\tool generated \textbf{114 correct rules} out of the 156 generated rules (\text{73.1\%}) from \textbf{auto-generated transition examples}. \textbf{96 (61.5\%) of those rules would not have been generated otherwise}.
\end{highlight}

\subsection{RQ3: Generalizability}

\newsavebox{\codeboxA}
\newsavebox{\codeboxB}
\newsavebox{\codeboxC}
\newsavebox{\codeboxD}
\newsavebox{\codeboxE}
\newsavebox{\codeboxF}
\newsavebox{\codeboxG}
\newsavebox{\codeboxH}
\newsavebox{\codeboxI}
\newsavebox{\codeboxJ}
\newsavebox{\codeboxK}
\newsavebox{\codeboxL}
\newsavebox{\codeboxQ}
\newsavebox{\codeboxW}
\newsavebox{\codeboxP}
\newsavebox{\codeboxU}

\begin{lrbox}{\codeboxA}
\begin{yourcodelisting}[language=langs, breaklines=true]
*:[[x]]*.set_index(*:[a]*,  drop=*:[[b]]*, inplace=True)
\end{yourcodelisting}
\end{lrbox}

\begin{lrbox}{\codeboxB}
\begin{yourcodelisting}[language=langs, breaklines=true]
*:[[x]]*.read_csv(*:[[a]]*, compression=*:[[b]]*,
 encoding=*:[[c]]*, index_col=*:[d]*, squeeze=True)
\end{yourcodelisting}
\end{lrbox}

\begin{lrbox}{\codeboxH}
\begin{yourcodelisting}[language=langs, breaklines=true]
jaccard_similarity_score(*:[[a]]*, *:[[b]]*)
\end{yourcodelisting}
\end{lrbox}

\begin{lrbox}{\codeboxC}
\begin{yourcodelisting}[language=langs, breaklines=true]
BaggingClassifier(base_estimator=*:[[a]]*, 
 n_estimators=*:[[b]]*, random_state=*:[[c]]*)
\end{yourcodelisting}
\end{lrbox}

\begin{lrbox}{\codeboxD}
\begin{yourcodelisting}[language=langs, breaklines=true]
BaggingRegressor(base_estimator=*:[[a]]*, 
 n_estimators=*:[[b]]*, random_state=*:[[c]]*)
\end{yourcodelisting}
\end{lrbox}

\begin{lrbox}{\codeboxE}
\begin{yourcodelisting}[language=langs, breaklines=true]
KMeans(n_clusters=*:[a]*, init=*:[[b]]*, 
 n_init=*:[[c]]*, algorithm='full')
\end{yourcodelisting}
\end{lrbox}

\begin{lrbox}{\codeboxF}
\begin{yourcodelisting}[language=langs, breaklines=true]
AgglomerativeClustering(n_clusters=*:[a]*, 
 linkage=*:[b]*, affinity=*:[c]*)
\end{yourcodelisting}
\end{lrbox}

\begin{lrbox}{\codeboxG}
\begin{yourcodelisting}[language=langs, breaklines=true]
OneHotEncoder(sparse=*:[[aac]]*, 
 categories=*:[[aan]]*, drop=*:[[aaz]]*)
\end{yourcodelisting}
\end{lrbox}

\begin{lrbox}{\codeboxJ}
\begin{yourcodelisting}[language=langs, breaklines=true]
*:[[x]]*.filters.gaussian_filter(*:[a]*, 
 *:[b]*, mode=*:[[c]]*)
\end{yourcodelisting}
\end{lrbox}

\begin{lrbox}{\codeboxK}
\begin{yourcodelisting}[language=langs, breaklines=true]
*:[[x]]*.query(*:[[a]]*, *:[[b]]*, n_jobs=*:[c]*)
\end{yourcodelisting}
\end{lrbox}

\begin{lrbox}{\codeboxL}
  \begin{yourcodelisting}[language=langs, breaklines=true]
*:[[x]]*.hanning(*:[[a]]*, *:[[b]]*)
  \end{yourcodelisting}
  \end{lrbox}

\begin{lrbox}{\codeboxQ}
  \begin{yourcodelisting}[language=langs, breaklines=true]
*:[[x]]*.alltrue(*:[a]*, axis=*:[b]*)
  \end{yourcodelisting}
  \end{lrbox}

\begin{lrbox}{\codeboxW}
\begin{yourcodelisting}[language=langs, breaklines=true]
*:[[x]]*.histogram(*:[[a]]*, bins=*:[b]*, range=*:[c]*, normed=*:[y]*)
\end{yourcodelisting}
\end{lrbox}

\begin{lrbox}{\codeboxP}
\begin{yourcodelisting}[language=langs, breaklines=true]
*:[[x]]*.complex(*:[[a]]*, *:[[b]]*)
\end{yourcodelisting}
\end{lrbox}

\begin{lrbox}{\codeboxU}
  \begin{yourcodelisting}[language=langs, breaklines=true]
*:[[aai]]*.apply(*:[a]*, axis=*:[[b]]*, reduce=True)
  \end{yourcodelisting}
  \end{lrbox}


\begin{table*}[t]
  \centering
  \caption{\small \textbf{RQ3.} Comparison of Non-General and Generalized Rules}
  \label{tab:generalizability}
  \resizebox{\textwidth}{!}{%
  \begin{tabular}{l lS lS}
  \toprule
  \textbf{Library} & \multicolumn{2}{c}{\textbf{Original Rule}} & \multicolumn{2}{c}{\textbf{Generalized Rule}}  \\
  \cmidrule(lr){2-3} \cmidrule(lr){4-5}
  & \textbf{Match Template} & \textbf{Matches} & \textbf{Match Template} & \textbf{Matches} \\
  \midrule
  \multirow{2}{*}[-2em]{\pandas}
  & \usebox{\codeboxA} & 2 & \lstinline[language=langs]|*:[[x]]*.set_index(*:[args]*, inplace=True)| & 370  \\
  \cmidrule(lr){2-5}
  & \usebox{\codeboxB} & 0 & \lstinline[language=langs]|*:[[x]]*.read_csv(*:[args]*, squeeze=True)| & 21   \\
  \cmidrule(lr){2-5}
  & \usebox{\codeboxU} & 3 & \lstinline[language=langs]|*:[[aai]]*.apply(*:[args]*, reduce=True)| & 4   \\
  \midrule
  \multirow{4}{*}[-1.6em]{\scipy}
  & \usebox{\codeboxH} & 94 & \lstinline[language=langs]|jaccard_similarity_score(*:[args]*)| & 226   \\
  \cmidrule(lr){2-5}
  & \usebox{\codeboxJ} & 0 & \lstinline[language=langs]|*:[[x]]*.filters.gaussian_filter(*:[args]*)| & 86   \\
  \cmidrule(lr){2-5}
  & \usebox{\codeboxK} & 0 & \lstinline[language=langs]|*:[[x]]*.query(*:[args]*, n_jobs=*:[y]*)| & 0   \\
  \cmidrule(lr){2-5}
  & \usebox{\codeboxL} & 0 & \lstinline[language=langs]|*:[[x]]*.hanning(*:[args]*)| & 0  \\
  \midrule
  \multirow{1}{*}[-1.6em]{\numpy}
  & \usebox{\codeboxQ} & 7 & \lstinline[language=langs]|*:[[x]]*.alltrue(*:[args]*)| & 208   \\
  \cmidrule(lr){2-5}
  & \usebox{\codeboxW} & 2 & \lstinline[language=langs]|*:[[x]]*.histogram(*:[args]*, normed=*:[y]*)| & 66   \\
  \cmidrule(lr){2-5}
  & \usebox{\codeboxP} & 17 & \lstinline[language=langs]|*:[[x]]*.complex(*:[args]*)| & 20   \\
    \midrule
   \multirow{6}{*}[-2.6em]{\sklearn}
  & \usebox{\codeboxC} & 26 & \lstinline[language=langs]|BaggingClassifier(base_estimator=*:[x]*, *:[args]*)| & 220   \\
  \cmidrule(lr){2-5}
  & \usebox{\codeboxD} & 7 & \lstinline[language=langs]|BaggingRegressor(base_estimator=*:[x]*, *:[args]*)| & 116   \\
  \cmidrule(lr){2-5}
  & \usebox{\codeboxE} & 0 & \lstinline[language=langs]|KMeans(*:[args]*, algorithm='full')| & 38   \\
  \cmidrule(lr){2-5}
  & \usebox{\codeboxF} & 4 & \lstinline[language=langs]|AgglomerativeClustering(*:[args]*, affinity=*:[c]*)| & 28   \\
  \cmidrule(lr){2-5}
  & \usebox{\codeboxG} & 0 & \lstinline[language=langs]|OneHotEncoder(sparse=*:[x]*, *:[args]*)| & 66   \\
  \bottomrule  
\end{tabular}
  }%
\end{table*}

Of the 156 rules we manually validated in RQ2, 41 had generalized arguments, and only 9 (22\%) were incorrect. 
To further evaluate the impact of generalizability, we conducted an ablation study by disabling the generalization procedure.
We selected 15 rules that had been generalized, along with their non-generalized counterparts. 
Using Sourcegraph's code search~\cite{sourcegraph}, we searched for repositories containing a given keyword in the rule (e.g., for \texttt{readcsv(..., squeeze=True)}, we searched for \texttt{squeeze=True}). 
We then cloned up to 50 random repositories for each rule, and ran the generalized and non-generalized rules on these repositories, counting matches. 

Table~\ref{tab:generalizability} compares the matches for original and generalized rules, showing that generalization significantly improves rules applicability. 
For instance, the number of matches for the \lstinline|set_index| case increased from 2 to 370 (185x) with generalization. Generalization is important because it abstracts context unrelated to the actual API change. 
This is particularly important in Python, where APIs can have many argument combinations (e.g., APIs taking 10 keyword arguments). Notice that some rules had zero matches. This happened because \comby could not infer types for template variables (\comby does not apply rules if it cannot infer types) or our heuristic query search on Sourcegraph did not find actual API usages.


\begin{highlight}
Generalization led to a \textbf{9.07x increase} in rule matches, boosting potential rule applications from \textbf{162} to \textbf{1469} in our sample. This demonstrates the \textbf{significant impact of generalization on rule applicability}.
\end{highlight}

\begin{table*}[t]
  \caption{\small \textbf{RQ4.} Effects of rule application on developer projects.}
  \label{tab:dev-projects}
  \small
  \centering
  \begin{tabular}{lrrrrrrrrr}
  \toprule
  \textbf{Library} &  \makecell{\textbf{Total} \\ \textbf{Projects}} & \makecell{\textbf{Affected} \\ \textbf{Projects}} & 
  \makecell{\textbf{Unique} \\ \textbf{Rules}} &
  \makecell{\textbf{Rule} \\ \textbf{Applications}} & \makecell{\textbf{Additional} \\ \textbf{Warnings}} & \makecell{\textbf{Resolved} \\ \textbf{Warnings}} & \makecell{\textbf{Additional} \\ \textbf{Passing Tests}} & \makecell{\textbf{Additional} \\ \textbf{Failures}} & \makecell{\textbf{Resolved} \\ \textbf{Failures}} \\ \midrule    
  \sklearn & 20 & 10 & 6 & 27 & 9 & 598 & 2 & 1 & 1 \\
  \pandas & 20 & 10 & 4 & 23 & 0 & 44 & 7 & 81 & 7  \\
  \scipy & 20 & 6 & 5 & 23 & 0 & 266 & 0 & 1 & 0 \\
  \midrule
  \textbf{Total} & 60 & 26 & 15 & 73 & 9 & 908 & 9 & 83 & 8 \\
  \bottomrule
  \end{tabular}
\end{table*}

\subsection{RQ4: Updating client code}

To evaluate the effectiveness of our approach to updating developer code, we migrated outdated library API usage in developer projects found on GitHub for the \sklearn, \pandas, and \scipy libraries. 
Collecting and running client projects requires significant manual effort: many projects do not specify dependencies or provide tests. We therefore did not evaluate \numpy API usage, but we can expect similar results.

We found client projects by searching GitHub for public repositories that used outdated versions of each library, and included code that matched to at least one of the match templates of an inferred rule from RQs 1 and 2.
We applied a total of 15 unique rules across the three libraries. We provide detail on specific rules and projects in Zenodo~\cite{melt_artifact}.
For each library, we identified 20 client projects that used outdated versions, and between one and three rules applied.  
We cloned each project, updated its library dependencies to a version with the breaking change, installed necessary dependencies, and ran all tests to note passing tests, failures, errors, and warnings.  We then used \comby to automatically update the outdated API usage, and reran the tests to compare results post-migration. We did this separately for each applicable rule.  

Table~\ref{tab:dev-projects} summarizes results. 
Total Projects refers to the total number of projects to which we applied rules and tested. Affected Projects refers to the number of evaluated projects that had a change in the tests after rule application from new or resolved warnings, passed tests, or failures. 
Not all of the projects had tests affected by rule application, either because test coverage was incomplete or because persistent failing tests in developer projects obscured the effect of rule application.

For \sklearn, half the developer project tests were affected by rule application. Only two of the projects showed a negative impact of rule application, where one project had an additional failing test and another project had nine new warnings. 
The \sklearn rules were applied without type information, which is one potential cause for the negative impact. 
The other affected projects had warnings resolved, ranging from 1 to 563 warnings resolved for a single project. One project had additional passing tests. 

For \pandas, rule application affected half of client projects. 
While there were 81 additional failures from \pandas rules, they were isolated to four projects and a single rule.
These new failures occurred due to lack of type information in one rule, which resulted in its erroneous application to API calls unrelated to the \pandas library.
In other projects, the same rule was applied correctly, even without type information, and successfully resolved warnings. The other three unique \pandas rules were applied with type information. 
No \pandas rules introduced new warnings.

For \scipy, the 5 rules were also had no type information, but only one application introduced an error. 
All six affected \scipy projects had warnings resolved by rule application, and none of the \scipy rule applications caused additional warnings. 

Of the 60 evaluation projects, 34 had no change in the tests or warnings.
However, this does not indicate that rule transformation was incorrect or unnecessary: most projects had failing tests and errors unrelated to API usage, which can obscure the effect of rule application. 
Overall, the resolved warnings and failures demonstrate \tool's potential to help developers more easily maintain large projects.


\newsavebox{\codeboxAA}
\newsavebox{\codeboxAB}
\newsavebox{\codeboxAC}
\newsavebox{\codeboxAD}
\newsavebox{\codeboxAE}
\newsavebox{\codeboxAF}
\newsavebox{\codeboxAG}
\newsavebox{\codeboxAH}
\newsavebox{\codeboxAI}
\newsavebox{\codeboxAJ}
\newsavebox{\codeboxAK}
\newsavebox{\codeboxAL}
\newsavebox{\codeboxAM}
\newsavebox{\codeboxAN}
\newsavebox{\codeboxAO}
\newsavebox{\codeboxAQ}
\newsavebox{\codeboxAR}
\newsavebox{\codeboxAP}

\begin{lrbox}{\codeboxAA}
  \begin{yourcodelisting}[language=langs, mathescape]
OneHotEncoder(sparse=*:[aao]*) $\rightarrow$ OneHotEncoder(sparse_output=*:[aao]*)
  \end{yourcodelisting}
\end{lrbox}

\begin{lrbox}{\codeboxAB}
  \begin{yourcodelisting}[language=langs]
OneHotEncoder(:[gen_args_0], sparse=False) $\rightarrow$ OneHotEncoder(:[gen_args_0], sparse_output=False)
  \end{yourcodelisting}
\end{lrbox}

\begin{lrbox}{\codeboxAC}
  \begin{yourcodelisting}[language=langs]
:[[aau]].append(:[[aas]]) $\rightarrow$ pd.concat([:[[aau]], :[[aas]]])
  \end{yourcodelisting}
\end{lrbox}

\begin{lrbox}{\codeboxAD}
  \begin{yourcodelisting}[language=langs]
:[[aai]].read_csv(:[gen_args_0], squeeze=True) $\rightarrow$ :[[aai]].read_csv(:[gen_args_0]).squeeze("columns")
  \end{yourcodelisting}
\end{lrbox}

\begin{lrbox}{\codeboxAE}
  \begin{yourcodelisting}[language=langs]
:[[aad]].set_index(:[gen_args_0], inplace=True) $\rightarrow$ :[[aad]] = :[[aad]].set_index(:[gen_args_0], copy=False)
  \end{yourcodelisting}
\end{lrbox}

\begin{lrbox}{\codeboxAF}
  \begin{yourcodelisting}[language=langs]
read_table(:[aae]) $\rightarrow$ read_csv(:[aae])
  \end{yourcodelisting}
\end{lrbox}

\begin{lrbox}{\codeboxAG}
  \begin{yourcodelisting}[language=langs]
eigh(:[gen_args_0], eigvals=:[aac], check_finite=:[aaf]) $\rightarrow$ eigh(:[gen_args_0], subset_by_index=:[aac], check_finite=:[aaf])
  \end{yourcodelisting}
\end{lrbox}

\begin{lrbox}{\codeboxAH}
  \begin{yourcodelisting}[language=langs]
:[[aae]].filters.gaussian_filter(:[gen_args_0]) $\rightarrow$ :[[aae]].gaussian_filter(:[gen_args_0])
  \end{yourcodelisting}
\end{lrbox}

\begin{lrbox}{\codeboxAI}
  \begin{yourcodelisting}[language=langs]
ndimage.filters.median_filter(:[gen_args_0]) $\rightarrow$ ndimage.median_filter(:[gen_args_0])
  \end{yourcodelisting}
\end{lrbox}

\begin{lrbox}{\codeboxAJ}
  \begin{yourcodelisting}[language=langs]
minimize(:[gen_args_0], options
\end{yourcodelisting}
\end{lrbox}

\begin{lrbox}{\codeboxAK}
  \begin{yourcodelisting}[language=langs]
has_fit_parameter(ensemble.base_estimator_, :[aat]) $\rightarrow$ has_fit_parameter(ensemble.estimator_, :[aat])
  \end{yourcodelisting}
\end{lrbox}

\begin{lrbox}{\codeboxAL}
\begin{yourcodelisting}[language=langs]
pinv2(:[[aao]]) $\rightarrow$ pinv(:[[aao]])
\end{yourcodelisting}
\end{lrbox}

\begin{lrbox}{\codeboxAM}
\begin{yourcodelisting}[language=langs]
KMeans(:[gen_args_0], algorithm='full') $\rightarrow$ KMeans(:[gen_args_0])
\end{yourcodelisting}
\end{lrbox}

\begin{lrbox}{\codeboxAN}
\begin{yourcodelisting}[language=langs]
safe_indexing(:[gen_args_0]) $\rightarrow$ safe_indexing(:[gen_args_0], axis=0)
\end{yourcodelisting}
\end{lrbox}

\begin{lrbox}{\codeboxAO}
\begin{yourcodelisting}[language=langs]
base_estimator.score(:[gen_args_0]) $\rightarrow$ estimator.score(:[gen_args_0])
\end{yourcodelisting}
\end{lrbox}

\section{Discussion}
\label{sec:discussion}

In this section, we address the main limitations of our method and possible future work.

\subsection{Limitations and threats}

\noindent\textbf{Rule correctness.}
We used manual validation to assess rule correctness, with a process that entailed high IRR kappa indicating agreement. 
One approach for further validation could involve upgrading client projects to newer library versions and applying the rules on projects using these libraries. In RQ4, we use this method to demonstrate that some rules are indeed correct. However, this process is challenging. \tool does not mine rules for \emph{all} breaking changes in a given release, so upgrading client projects may break multiple aspects in ways automatic find-and-replace rules cannot address~\cite{impossible_migration}. 
However, automating a large part of migration in ways that entail minimal additional technology or effort on the part of the client developer holds promise for reducing the challenge of upgrading library dependencies. 

\vspace{1em}
\noindent\textbf{Code generation model.}
Our approach relies on a code generation model to generate examples when none are available. We selected \texttt{gpt-4-0314}, a state-of-the-art model trained on data before September 2021.
We successfully evaluated our approach on PRs opened after September 2021, demonstrating that the risk of data leakage in these experiments is low.
The model, however, is paid and not open-source. 
As AI research advances, we anticipate better models being made public. 
We opt for a model-based code generation approach over generating \comby rules directly with the model because:
(1) rules can be validated with code examples (if the code does not pass, we discard the example), (2) the model is not fine-tuned, has limited exposure to \comby, and is likely to work better on commonly-used languages like Python.
For less popular APIs, however, fine-tuned versions of the model on library code might be necessary. 

\vspace{1em}
\noindent\textbf{Generalization.}
Our generalization procedure removes context and arguments that appear unrelated to the change. Removing too much context and type information may result in spurious rules. Conversely, insufficient generalization can make the rule too specific. \tool can return both rules to the user, allowing them to decide what to keep. Currently, developers must manually validate rules to ensure they make sense. To facilitate this, we developed a CI solution on GitHub for integrating our tool. Rules can be validated and modified, if necessary, by whoever merges the PR, or automatically validated, as previously discussed.

\subsection{Comparison against prior work}
\label{sec:compare}

Few API migration tools target Python, challenging direct comparison to prior work. \tool adapts its inference algorithm from \inferrule~\cite{inferrule}, designed for type migration in Java. 
Consequently, \tool without generalization and filtering serves as a baseline equivalent to \inferrule. 
The most closely related approach, \PyEvolve~\cite{pyevolve}, builds on \inferrule using \comby as an intermediate representation. 
\PyEvolve focuses on general refactoring,  and adapts rules to different control variants, requiring more complex analysis, and client code analysis. 
 This is in contrast to \tool's lightweight approach, which aims to minimize overhead on client developers.
Since most of our rules are 1:1 and 1:n transformations, adapting rules for control flow variants is less relevant. Overall, while \PyEvolve is more powerful in the types of rules it can infer, fundamentally it serves a different goal as compared to \tool.

Our evaluation differs from closely-related prior work~\cite{meditor, apifix} in two ways.  First, our manual validation process is able to consider more information in the form of the PR and library documentation. %
That is, rather than looking at rules in isolation or limiting attention to syntactic validity, we can consider whether the change actually reflects PR intent.  Second, we provide an end-to-end evaluation of automatically inferred rules on a number of client code repositories, complementing manual rule validation.  

As we discuss in Section~\ref{sec:related}, most prior approaches for automatic
API migration (or code evolution generally) mine migration examples from client
projects or their source control histories. \tool relies solely on the changed
\emph{library}, looking at internal code changes to inform rule mining. This
allows \tool to apply earlier in the library update process. However, libraries
do not always include sufficient changed code examples to inform migration,
which is why \tool also prompts an LLM to generate extra examples, along with
tests to validate those examples. Other approaches may also benefit from using LLMs this
way, particularly those whose use cases entail fewer available examples, like
\texttt{A3}~\cite{a3} (focusing on Android API migration), or \texttt{APIFix}~\cite{apifix}
(evaluated on changes to library code, similar to \tool). \texttt{APIFix} in particular
could likely benefit from the LLM-generated examples and tests, because it uses edit examples in its program synthesis algorithm.
Other tools are evaluated across many more
example changes to client code, like \texttt{Meditor}~\cite{meditor}. These approaches
may not require new examples, but
leveraging LLMs may allow them to apply earlier in the update process, or in scenarios where migration examples are scarce.
Indeed, as models with larger context windows become available (e.g., CLAUDE 100K
token context \cite{claude_100k_context_windows}), it becomes 
possible to include more comprehensive data in prompts, such as full API
documentation. This suggests a promising avenue for generating higher-quality,
context-rich examples for rule mining, particularly when extant migration
examples are scarce.

\section{Related Work}
\label{sec:related}

\noindent\textbf{Empirical studies on API evolution.}
API evolution has long been a challenging software engineering concern without a definitive solution. 
Developers often lag in updating their software to the latest APIs, leading to compatibility issues and hindering maintenance~\cite{mcdonnell2013empirical}. 
Recent work identifies a significant need for more support for API evolution tools in languages other than Java, particularly Python~\cite{api_evolution_review}. 
\tool aims to address this gap. Moreover, Dilhara et al. \cite{ml_needs_help} further reinforce the need for migration tools for Python, finding that Python data library clients tend to need to update dependencies frequently and face significant challenges in doing so.

A study of API migration in four popular Java libraries found that only 27.1\% of these migrations were fully automatable~\cite{impossible_migration}. 
This suggests that achieving 100\% safe and automated migration rules is unlikely, as some transformations are complex and need more context than rule-based approaches can provide. 
\tool's imperfect accuracy aligns with these findings, as some rules are incorrect simply due to the difficulty in capturing them with purely syntax-driven transformations. However, \tool's approach can provide semi-automated support for migration, easing the overall burden.


Meanwhile, refactoring tools that require developers to use complex domain-specific languages are often difficult to use and, consequently, often poorly-adopted~\cite{dsl_are_bad}. 
This observation emphasizes the need to develop user-friendly and ergonomic API migration tools and techniques that seamlessly integrate into developers' workflows, as \tool aims to do.



\vspace{1em}
\noindent\textbf{Library evolution.}
Automated API migration research has primarily focused on mining client repositories, usually targeting object-oriented languages (namely Java and C\#).  For example, \texttt{A3} \cite{a3} and \texttt{Meditor} \cite{meditor} mine client repositories for examples to create rules, which are then customized to new clients. \texttt{APIFix} \cite{apifix} mines transition examples from both previously-migrated and new client repositories and uses \texttt{Refazer}'s \cite{refazer} engine to learn rules. \texttt{Refazer}'s transformations are expressed as AST edits, which are more difficult to understand \cite{recode}. Unlike previous approaches, \tool emphasizes simplicity via lightweight find-and-replace transformations.


\texttt{APIMigrator}~\cite{apimigrator} and \texttt{AppEvolve}~\cite{appevolve} also mine client repositories for transition examples and apply them directly to clients. Both tools use differential testing to validate edits on clients. \tool focuses on generating rules rather than updating client code directly, however, it could similarly benefit from incorporating differential testing \cite{differential_testing} to validate inferred rules.

Approaches like \texttt{Semdiff}~\cite{semdiff} recommends API changes to developers by presenting a ranking of potential replacements. 

\vspace{1em}
\noindent\textbf{Code Refactoring.}
\texttt{Catchup!} \cite{catchup} records refactorings made by library developers during development and replaces them in client code. \texttt{LASE} \cite{lase} and \texttt{SYEDIT} \cite{syedit} mine code examples for systematic edits, generating edit scripts at the AST level rather than using find-and-replace rules. \inferrule \cite{inferrule, tcinfer} inspired our rule inference algorithm. However, \inferrule primarily targets type migration and extensively mines client repositories for refactoring examples.


\PyEvolve \cite{pyevolve}, developed concurrently with this work, also uses \inferrule's algorithm to infer \comby rules from code changes. However, its purpose is different, as we discussed in detail in Section~\ref{sec:compare}.
Most API migrations are either 1:1 (47.2\%) or 1:n (48.1\%)~\cite{impossible_migration}, and control-flow awareness is not necessary for API evolution. \tool focuses on rule generalization instead. This approach reduces the analysis overhead introduced by \PyEvolve, especially for large codebases, since only \comby needs to be run.

\SOAR~\cite{soar} uses program synthesis to refactor client code rather than generate find-and-replace rules, aiming to support migration \emph{between} libraries. Unlike \tool, \SOAR refactors client code as a blackbox, which can be less interpretable.
It can handle more complex migrations, but is commensurately less performant.
To the best of our knowledge, \PyEvolve and \SOAR are the only two tools besides \tool that can infer and apply refactorings for Python code. 


\section{Conclusion}

Selecting and maintaining APIs is critical yet challenging in software development. 
Developers may have to manually update APIs due to evolving libraries, which is time-consuming and error-prone process. 
We present \tool, which assists developers by generating lightweight API Migration rules in \comby. 
Unlike previous approaches, \tool mines rules directly from library pull requests instead of client projects.
This approach allows rule inference to be integrated directly into the library workflow, eliminating the need to wait for clients to migrate their code. Furthermore, \tool rules are purely syntax-driven, and require no additional tooling on client side (besides \comby). 
We evaluated \tool on pull requests from four popular libraries: \pandas, \scipy, \numpy, and \sklearn. 
We assessed rule accuracy by examining the pull request descriptions, discussions, and more. 
We discovered 461 accurate rules from code examples in pull requests and 114 rules from auto-generated code examples. 
To show practical applicability, we applied the rules to client projects and ran their tests, proving their effectiveness in real-world situations.

\section*{Acknowledgements}
This work was supported by Portuguese national funds through FCT under projects UIDB/50021/2020, PTDC/CCI-COM/2156/2021, 2022.03537.PTDC and grant SFRH/BD/150688/2020, as well as the US National Science Foundation, Awards CCF-1750116 and CCF-1762363.

\bibliographystyle{IEEEtran}
\bibliography{refs}
\end{document}